\documentclass[prl,twocolumn,superscriptaddress,showpacs]{revtex4}
\usepackage{epsfig,amsmath,amssymb,graphics,color,calc}
\newcommand{\be}{\begin{equation}}
\newcommand{\ee}{\end{equation}}
\newcommand{\ba}{\begin{eqnarray}}
\newcommand{\ea}{\end{eqnarray}}
\newcommand{\T}{^{\mathrm{T}}}
\newcommand{\B}{_{\mathrm{B}}}

\begin{document} 
\title{Suppressed compressibility at large scale 
in jammed packings of size disperse spheres}

\author{Ludovic Berthier}
\affiliation{Laboratoire Charles Coulomb, 
UMR 5221 CNRS and Universit\'e Montpellier 2, Montpellier, France}

\author{Pinaki Chaudhuri}
\affiliation{Laboratoire PMCN, Universit\'e Lyon 1,
Universit\'e de Lyon, UMR CNRS 5586, 69622 Villeurbanne, France}

\author{Corentin Coulais}
\affiliation{Service de Physique de l'Etat Condens\'e, 
CEA-Saclay, URA 2464 CNRS, 91191 Gif-sur-Yvette, France}

\author{Olivier Dauchot}
\affiliation{Service de Physique de l'Etat Condens\'e, 
CEA-Saclay, URA 2464 CNRS, 91191 Gif-sur-Yvette, France}

\author{Peter Sollich}
\affiliation{King's College London, Department of Mathematics, Strand,
London WC2R 2LS, U.K.}

\date{\today}
\begin{abstract}
We analyze the large scale structure and fluctuations of jammed packings
of size disperse spheres, produced in a granular experiment as well as
numerically.
While the structure factor of the packings reveals no 
unusual behavior for small wavevectors, the compressibility displays 
an anomalous linear dependence at low
wavectors and vanishes when $q \to 0$. 
We show that such behavior occurs because jammed packings of
size disperse spheres have no bulk fluctuations of the volume fraction 
and are thus
hyperuniform, a property not observed experimentally before.
Our results apply to arbitrary particle size 
distributions. For continuous distributions, we 
derive a perturbative expression for the compressibility 
that is accurate for polydispersity up to about 30\%.
\end{abstract}



\maketitle

When an assembly of hard particles is compressed, there comes a 
point where further compression is difficult 
because the required pressure is too large. A 
similar ``jammed'' state can be obtained
with soft repulsive particles (as in emulsions or foams), 
at a particular volume fraction $\phi_c$ above which the 
stability of the packing is controlled by the 
elasticity of the particles. A large body of recent work
in both theory and experiment has characterized the properties 
of jammed packings~\cite{ohern,vanhecke}. 

While much attention has focussed on contacts
at the interparticle scale, and force networks 
and connectivity at larger scales~\cite{ohern,vanhecke,vanhecke2},
there has been comparatively little research into fluctuations and response 
at very large scales. In Ref.~\cite{donev} the low-wavevector behaviour of 
the structure factor, $S(q)$,
in a monodisperse system of hard frictionless spheres 
was studied numerically, revealing ``unexpectedly'' weak density 
fluctuations at low $q$,
\be 
S(q) \approx  a q,
\label{linear}
\ee
for some constant $a$, as was also found later
 for soft particles~\cite{silbert}. 
This should be compared to 
the behaviour in liquids
(including hard sphere fluids at $\phi < \phi_c$), where
$S(q) \approx S(q \to 0) + a' 
q^2$~\cite{hansen}. 
These results imply that
bulk fluctuations in the number density are suppressed
at $\phi_c$: in $d$ dimensions, 
fluctuations of particle number in subsystems
of linear size $L$ scale as $\left\langle \Delta N^2 
\right\rangle_L \sim L^{d-1}$.
Such suppressed density fluctuations are the defining
feature of ``hyperuniform'' materials~\cite{torquato1}.

Two recent papers reported surprising 
results, failing to detect the behaviour 
in Eq.~(\ref{linear}). A numerical simulation of binary mixtures~\cite{xu}
found that (\ref{linear}) only holds for the particular case of 
a monodisperse system. Similarly, a 
confocal microscopy study~\cite{weeks} 
of a jammed system of moderately
polydisperse colloidal hard spheres also 
failed to observe Eq.~(\ref{linear}). 
Both studies report that $S(q)$ in size disperse systems is different from 
Eq.~(\ref{linear}), and suggest that size disperse packings might not 
be hyperuniform.
That this is a highly topical question is demonstrated by the recent
work of Ref.~\cite{arxiv}, where independently of our approach
hyperuniformity was detected, using two-point probability
functions.

Here, we report the first experimental observation of vanishing 
fluctuations of the
volume fraction and hyperuniformity in a granular experiment. 
The same observation is made for numerically produced polydisperse packings
with arbitrary size distributions.
In contrast with 
\cite{weeks,xu} we consider not only the structure factor
$S(q)$, but also the isothermal compressibility, $\chi_T(q)$, 
the latter being central to our analysis of 
size disperse packings. We propose a novel perturbative approach for extracting
$\chi_T(q)$ for continuous size distributions, which in general is 
a non-trivial task, and explain the connection
between vanishing compressibility and hyperuniformity.

First, we briefly describe our systems.
Experimentally we produce dense random granular packings by slowly
compressing horizontally vibrated bidisperse brass disks. Typically
4500 large disks (diameter $5 \pm 0.025\,$mm) and 3500 small
disks (diameter $4 \pm 0.025\,$mm), surrounded by rigid
walls, are placed on an oscillating glass plate (amplitude $5\,$mm,
frequency $10\,$Hz).
The packing fraction is increased logarithmically slowly ($d\phi/d
\log(t) \simeq 10^{-2}$) until the force $F$ measured on the
compressing wall increases sharply from $F=0.05\,Mg$ to $F>Mg$, where
$M=2$~kg is the typical total
mass of the ``grains''. At that point the packing 
jams, grains stop moving and the force measured at the wall
remains finite when the vibration is switched off
(see Ref.~\cite{Lechenault} for more details). We
take a high resolution picture of the entire packing ($2048 \times 
2048$
pixels) at the largest packing fraction, correcting for optical distortion.
We detect the positions of the grains 
with a resolution of $20 \, \mu$m and 
retain $\approx 4000$ large grains and $\approx 
3000$ small grains at
least $1\,$cm away from the walls. 
We analyze two independent packings, 
produced starting from uncorrelated initial conditions. Rattler particles are 
always included in  the analysis.
Numerically we generate $3d$
polydisperse sphere packings at $\phi_c$
using soft repulsive particles, as first
proposed in \cite{ohern}, using conjugate gradient methods 
and small decompression steps to prepare 
packings exactly at $\phi_c$~\cite{ohern,pinaki}.
For each set of parameters, we prepare a single, very large
configuration composed 
of $N=64,000$ particles.
We study both a 50:50 binary mixture
of spheres with diameter ratio $R \in [1,2]$,
or systems with a continuous size distribution, which we take as 
a flat distribution centered around the average value $\bar\sigma$.
We studied polydispersities up to $p=0.4$, 
where $p$ is the standard deviation of the
size distribution divided by $\bar\sigma$.

We now provide some definitions.
Consider a size disperse system composed of 
$n$ species, containing $N_i$ particles of species $i$, with diameter
$\sigma_i$. We define 
$N = \sum_{i=1}^n N_i$, the density $\rho = N/V$, 
the concentrations
$x_i = N_i/N$, and the partial density fields
$\rho_i({\bf q}) = \sum_{j=1}^{N_i} e^{i {\bf q} \cdot {\bf r}_j}$, 
where ${\bf r}_j$ is the position of particle $j$ belonging  
to species $i$. 
The partial structure factors read
$S_{ij}(q) = \frac{1}{N} \left\langle \rho_i({\bf q}) \rho_j(-{\bf q})
\right\rangle$,
and we collect them in a matrix, ${\bf S}(q)$.
The total structure factor is defined as usual: 
\be
S(q) = \sum_{i=1}^n \sum_{j=1}^n S_{ij}(q). 
\label{eq-sq}
\ee

\begin{figure}
\psfig{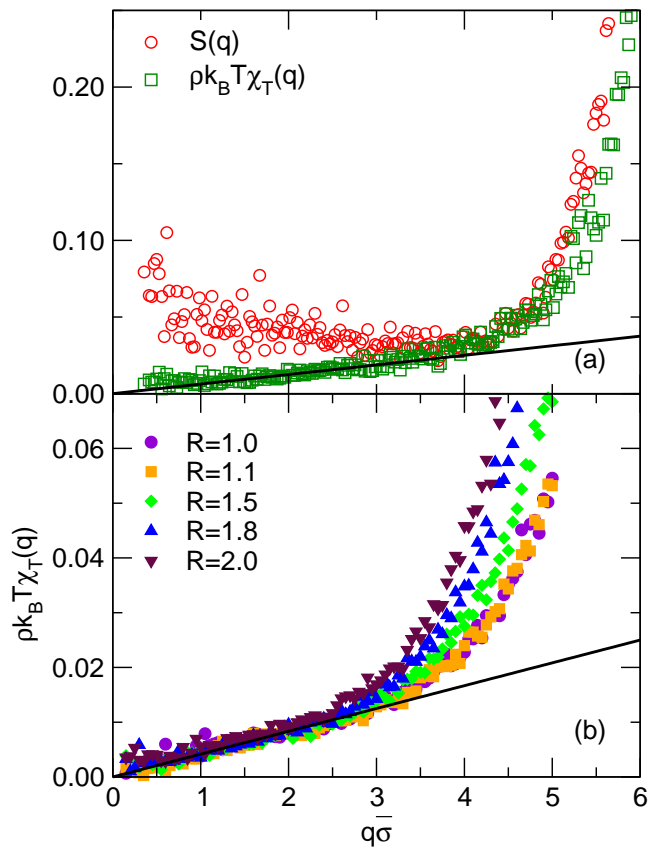}
\caption{\label{manip} (a) 
Structure factor and compressibility for the 
two-dimensional packings of disks obtained experimentally. While 
$S(q)$ resembles that of a binary fluid mixture, the compressibility
displays anomalous low-$q$ linear decay.
(b) Compressibility for 
numerically generated jammed packings of
50:50 binary mixtures with various size ratios $R$.
All systems share the same low-$q$ linear behaviour
of the compressibility.}
\end{figure}

When we analyze $S(q)$ in our experimental 
granular packings, see Fig.~\ref{manip}a,
we find that the low-$q$ behaviour is not compatible
with Eq.~(\ref{linear}). 
The same observation holds, see Fig.~\ref{eq-sq}a, 
for the evolution of $S(q)$ 
for numerical packings with continuous
size distribution of increasing polydispersity. 
The inset shows that $S(q \to 0)$
increases continuously with the polydispersity $p$.
This suggests that size dispersity is the main factor 
responsible for the numerical results in Ref.~\cite{xu} and the 
experimental ones in Ref.~\cite{weeks}, 
where $p \approx 0.05$ and $S(0) \approx 0.05$.

\begin{figure}
\psfig{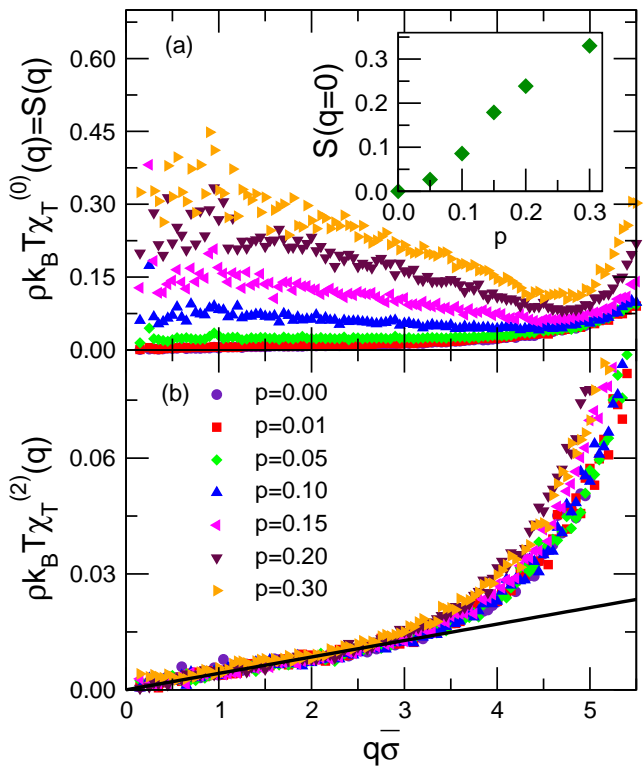}
\caption{\label{sq} (a) Zeroth and (b) second order 
estimate of the compressibility, Eq.~(\ref{ordera}), for packings
with continuous size distribution of polydispersity $p$. While the low-$q$
behaviour of the structure factor for $a=0$ is featureless
(see inset), 
the compressibility displays an anomalous linear 
behaviour at low-$q$, as seen for $a=2$. The second 
order estimate of the compressibility can be used 
up to $p \approx 30\%$, while the first order one (data not shown) becomes unreliable beyond $p\approx 10\%$.}
\end{figure}

For size disperse systems, we consider not only on $S(q)$ but also
the compressibility, $\chi_T(q)$. In matrix notation this reads, 
\be 
[\rho k\B T \chi_T(q)]^{-1}
= {\bf x}\T {\bf S}^{-1}(q) {\bf x},
\label{defchi}
\ee
where ${\bf x}\T = (x_1, \ldots, x_n)$.
To find $\chi_T$ one should 
measure all partial structure factors in 
${\bf S}(q)$, and invert this matrix to get ${\bf S}^{-1}(q)$.
For a one-component (monodisperse) system,
one finds $\rho k_B T \chi_T(q) = S(q)$, and both 
quantities are thus fully equivalent. 
For a binary mixture, $n=2$, one gets~\cite{batthia} 
\be
\rho k_B T \chi_T(q) = \frac{S_{11}(q)S_{22}(q) - S_{12}^2(q)}{  
x_1^2 S_{22}(q) + x_2^2 S_{11}(q) - 2 x_1 x_2 S_{12}(q) }.
\label{binary}
\ee
In Fig.~\ref{manip}a we show the compressibility 
measured experimentally. A clear linear behaviour of the
compressibility is obtained for low wavevectors.
To our knowledge, this anomalous behaviour has not been observed
experimentally before. 
In Fig.~\ref{manip}b we show the compressibility obtained numerically,
using Eq.~(\ref{binary}),
for binary mixtures with different size ratio.
For all systems considered, a linear behaviour of the
compressibility is obtained for low wavevectors.
This set of results suggests
that a relevant generalization of Eq.~(\ref{linear})
for binary mixtures is obtained by studying
$\chi_T(q)$, rather than $S(q)$.

While straightforward for discrete mixtures with a small number $n$ of
components, where 
${\bf S}$ is an $n \times n$ matrix, the matrix 
inversion in Eq.~(\ref{defchi}) is conceptually and
computationally difficult for continuous size distributions
where the size of the ${\bf S}$-matrix formally becomes infinite.
The system studied experimentally in \cite{weeks} is of this type.
To analyze such packings,
we derive a systematic approximation for $\chi_T(q)$.
The idea is that 
if the size distribution is sufficiently narrow, $\chi_T(q)$ can 
be obtained perturbatively. To this end, we define $\epsilon_i =
(\sigma_ i - \bar\sigma)/\bar\sigma$ and derive an expansion of $\chi_T(q)$
in powers of the $\epsilon_i$ up to some fixed but
otherwise arbitrary order $a$. To find a suitable starting point for
this expansion, recall the relation $S_{ij}(q) = x_i \delta_{ij} +
\rho x_i x_j h_{ij}(q)$ between the partial structure factors and
$h_{ij}(q)$, the Fourier transforms of the pair correlation functions
$g_{ij}(r)-1$~\cite{hansen}. The concentration factors $x_i$ vary
rapidly with $\epsilon_i$ for narrow distributions, so we cannot
expand $S_{ij}(q)$. But the pair correlation functions $g_{ij}(r)$ and
hence the $h_{ij}(q)$ depend smoothly on particle size, so we
expand the latter up to
$\epsilon_i^a \epsilon_j^a$, taking e.g.\ for $a=1$,
$h_{ij}(q)=h_0(q)+h_1(q)(\epsilon_i+\epsilon_j)+h_2(q)\epsilon_i 
\epsilon_j$~\cite{perturb}. This expansion is 
inserted into Eq.~(\ref{defchi}) for $S_{ij}(q)$, giving after some algebra
the compact form
\be
[\rho k\B T \chi_T^{(a)}(q)]^{-1} = 
{\bf m}_a\T {\bf S}_a^{-1}(q) {\bf m}_a,
\label{ordera}
\ee
where the matrix ${\bf S}_a(q)$ has elements 
$
S^{\mu \nu} (q) = \frac{1}{N} \left\langle \epsilon^\mu({\bf q}) 
\epsilon^\nu (-{\bf q}) 
\right\rangle, 
$
with $\mu,\nu \in \{0, \ldots, a \}$, and
captures fluctuations of the moment density fields
$\epsilon^\mu({\bf q})  =  
\sum_{i=1}^n  \epsilon_i^{\mu} \rho_i({\bf q})$.
Hence, $\epsilon^0({\bf q})=\rho({\bf q})$ is the number density
field, and $S^{00}(q)=S(q)$, the total structure factor.
The vector ${\bf m}_a^{\mathrm{T}} = (\delta_0, \ldots, \delta_a)$
in Eq.~(\ref{ordera}) has components given by the moments
of $\epsilon_i$ averaged over the 
particle size distribution:
$\delta_\mu = \sum_{i=1}^n x_i \epsilon_i^{\mu}$, so that 
$\delta_0=1$, $\delta_1=0$, and $\delta_2$ is directly related to 
the polydispersity, $\delta_2 = p^2$.
The result (\ref{ordera}) relates the compressibility to the 
$S^{\mu \nu}(q)$ up to order $a$. For $a$ not too large it is simple to
compute as it only requires the measurement of $(a+1)(a+2)/2$
reduced structure factors. It can be applied to 
arbitrary particle size distributions and is exact 
for discrete $n$-component
mixtures if we choose $a=n-1$, 
as can be shown by direct calculation from Eq.~(\ref{defchi}).

We have tested our general formula (\ref{ordera})
using computer simulations. When $a=0$, one has 
$\rho k_B T \chi^{(0)} = S(q)$, which is only exact for $n=1$
(monodisperse systems), as discussed above.
At first order, $a=1$, we need to invert a $(2 \times 2)$ matrix 
to get $\rho k\B T \chi_T^{(1)}(q)$. 
When applied to the case of a continuous size distribution 
this formula produces the expected linear behaviour at low $q$
for $p \leq 0.10$, 
but deviations appear at larger $p$.
To check whether these deviations are physical, or a result
of our approximation, we go to second order, $a=2$, 
where the required inversion of a $(3 \times 3)$ matrix gives:
\begin{widetext}
\be 
\rho k\B T \chi_T^{(2)}(q) =  
\frac{S^{00} S^{11} S^{22} + 2 S^{02} S^{01} S^{12} - S^{00} [S^{12}]^2 
- [S^{01}]^2 S^{22} - [S^{02}]^2 S^{11} }{ S^{11} S^{22} - 
[S^{12}]^2  + 2 \delta_2 ( S^{01} S^{12} - 
S^{02} S^{11} )  
+ \delta_2^2 (S^{00} S^{11} - [S^{01}]^2)    }.  
\ee 
\end{widetext}
We now find, see Fig.~\ref{sq}b, that a linear $q$-dependence is obtained for 
polydispersities as large as 30\%.  This suggests that the same behavior
should in fact be obtained for arbitrary size distributions, although 
measuring the compressibility is more difficult when $p$ is 
very large because we need to go to even higher orders $a$.

We now discuss the physical significance of our results.
It is perhaps not surprising, with hindsight, that
jammed sphere packings have vanishing compressibility since this is
precisely how the jamming transition was described in the opening lines
of the paper. However, the quantity we call 
``compressibility'' in this work is in fact a particular combination
of density fluctuations, Eq.~(\ref{defchi}), that only reduces 
to the compressibility at thermal equilibrium when the 
fluctuation-dissipation theorem holds~\cite{hansen}. 
Remarkably, our results suggest that a similar connection between
response and fluctuations may exist far from equilibrium 
near jamming, even in real granular packings. 
An obvious connection between response and fluctuations 
holds at $q=0$, since $\chi_T(0) = \phi^{-1} \partial \phi / \partial P$ 
indeed vanishes near jamming where $P \sim (\phi_c - \phi)^{-1}$, so that
both sides of Eq.~(\ref{defchi}) vanish. 
It would be interesting to extend these considerations to 
finite $q$ near the jamming transition.

Our results also illustrate that, for size disperse systems, the limit
$S(q \to 0)$ is in general 
not directly related to the isothermal compressibility, 
$\chi_T(0)$~\cite{scattering}. 
While the latter vanishes in jammed packings, 
the former is free to take any positive value. 
This explains why previous work on size disperse
packings failed to observe any anomalous
behaviour~\cite{weeks,xu}. Both quantities
are related to the amplitude of fluctuations of the number density,
but $S(q \to 0) \sim \left\langle \Delta N^2 \right\rangle$ 
captures the total fluctuation of $N$
while $\chi_T(q\to0)$ quantifies the fluctuations of $N$ {\it at fixed
composition}. 
This can be seen by defining the
composition fluctuation fields $c_i({\bf q}) = \rho_i({\bf q}) - x_i
\rho({\bf q})$ for $i=1,\ldots,n-1$. The compressibility from
Eq.~(\ref{defchi}) can then be rewritten as
\be 
\rho k\B T \chi_T(q) =  S(q) - {\bf s}_{0c}\T(q){\bf S}_{cc}^{-1}(q)
{\bf s}_{0c}(q)
\label{eq:hill}
\ee
where the $(n-1)$-dimensional vector ${\bf s}_{0c}$ gathers the
correlations between number and 
composition fluctuations, and the matrix ${\bf S}_{cc}$ the correlations among
the latter~\cite{hill,batthia}.
For the compressibility to vanish at jamming the two terms
must cancel, which implies that local fluctuations
in $N$ become fully correlated with composition fluctuations. 
On the other hand, $S(q\to 0)$ remains 
positive because a local fluctuation of $N$ can be induced by 
a local fluctuation of the mixture composition:
fluctuations of $N$ do occur in jammed 
{\it size disperse} packings.
 
The behaviour of $S(q)$ in Fig.~\ref{sq} is clearly inconsistent
with Eq.~(\ref{linear}), as noticed previously~\cite{weeks,xu}. 
Does this imply that jammed
size disperse packings are not hyperuniform, as suggested in 
\cite{xu}? For point particles,
hyperuniformity refers to vanishing bulk 
fluctuations of the number density, as described in the introduction.
However, for an assembly of spherical particles  
hyperuniformity requires vanishing bulk fluctuations 
of the local volume fraction~\cite{torquato}.
While $N$ and $\phi$ are directly proportional for monodisperse
spheres like those studied in Ref.~\cite{donev}, 
they are not when the packing is size disperse, and thus 
no conclusion can be drawn from $S(q)$ alone. 

A connection between the anomalous compressibility studied 
in this work and suppressed fluctuations of the
volume fraction in hyperuniform packings can be established.
>From Eq.~(\ref{defchi}), we realize that whenever the ${\bf S}(q)$
matrix possesses at least one eigenvalue that goes 
to zero at low $q$, the compressibility vanishes.
We have diagonalized ${\bf S}(q)$ or ${\bf S}_a(q)$ as obtained 
in our numerical 
and experimental packings, and indeed found that 
in each case anomalous behaviour of the compressibility originates
from a single vanishing eigenvalue. This implies that 
there exists a particular linear combination of the
partial density fields which has no bulk fluctuations. A detailed
analysis of the corresponding eigenvectors shows that they are 
fully compatible with the local definition of the volume fraction, 
$\phi({\bf q}) = (\pi/6)\sum_{i=1}^n \sigma_i^d \rho_i({\bf q})$. 
This identification holds exactly for binary mixtures in our $3d$ 
simulations and in the 
$2d$ experimental packings. 
It also holds true for the packings with continuous size
distributions, to the same order in $\epsilon_i$ that we analyse for the
compressibility ($a=2$, which is accurate up to polydispersity $p=30$\%).
Indeed, we have checked that direct measurements 
of $I(q) = \left\langle \phi({\bf q}) 
\phi(-{\bf q}) \right\rangle$ coincide 
with the compressibility shown throughout this article in the low-$q$
regime where linear behaviour is observed. 
Therefore, the anomalous behaviour of the 
compressibility reveals the absence of bulk fluctuations of the 
volume fraction. This lack of fluctuations was detected also in the recent, independent, study of Ref.~\cite{arxiv}, using a rather different methodology. We conclude that all our size disperse jammed packings 
are hyperuniform. 

We have demonstrated anomalous behaviour of the
compressibility in jammed
size disperse packings of spheres both in simulations and
in a granular experiment, using for the case of continuous size 
distributions an efficient perturbative approach.
We have related this to suppressed
bulk fluctuations of the volume fraction, or hyperuniformity (see also~\cite{arxiv}, thus
revealing a structural signature of jamming not seen in the
conventional structure factor. Our work also raises intriguing questions
about the applicability of fluctuation-dissipation relations to
jammed systems.

We acknowledge financial support from ANR grants Syscom (PC) and 
Dynhet (LB, CC, OD), and R\'egion Languedoc-Roussillon (LB).



\begin{thebibliography}{99}

\bibitem{ohern}
C. S. O'Hern, S. A. Langer, A. J. Liu, and S. R. Nagel, 
Phys. Rev. Lett. {\bf 88}, 075507 (2002).

\bibitem{vanhecke} M. van Hecke,
J. Phys. Condens. Matter {\bf 22}, 033101 (2010).

\bibitem{vanhecke2} W. G. Ellenbroek, E. Somfai, M. van Hecke, and 
W. van Saarloos,
Phys. Rev. Lett. {\bf 97}, 258001 (2006).

\bibitem{donev} A. Donev, F. H. Stillinger, 
and S. Torquato, Phys. Rev. Lett. {\bf 95}, 090604 (2005).

\bibitem{silbert} 
L. E. Silbert and M. Silbert,
Phys. Rev. E {\bf 80}, 041304 (2009).

\bibitem{hansen}
J. P. Hansen and I. R. McDonald, {\it Theory of Simple Liquids}
(Elsevier, Amsterdam, 1986).

\bibitem{perturb}
P. Sollich, Phys. Rev. Lett. {\bf 100}, 035701 (2008).

\bibitem{torquato1}
S. Torquato and F. H. Stillinger, Phys. Rev. E {\bf 68}, 041113 (2003).

\bibitem{xu}
N. Xu and E. S. C. Ching, Soft Matter {\bf 6}, 2944 (2010).

\bibitem{weeks}
R. Kurita and E. R. Weeks, 
Phys. Rev. E {\bf 82}, 011403 (2010). 

\bibitem{arxiv}
C. E. Zachary, Y. Jiao, and S. Torquato arXiv:1008.2548.

\bibitem{pinaki} P. Chaudhuri, L. Berthier, and S. Sastry,
Phys. Rev. Lett. {\bf 104}, 165701 (2010). 

\bibitem{Lechenault} F. Lechenault, O. Dauchot, G.
Biroli, J. P. Bouchaud, Europhys. Lett. {\bf 83}, 46003 (2008).

\bibitem{batthia}
A. B. Bhatia and D. E. Thornton, Phys. Rev. B {\bf 2}, 3004 (1970).

\bibitem{scattering}
P. Salgi and R. Rajagopalan, 
Adv. Colloids Interf. Sci. {\bf 43}, 169 (1993).

\bibitem{hill} T. L. Hill, {\it Statistical mechanics, Principles
and selected applications}  
(Dover, New York, 1987).

\bibitem{torquato}
C. E. Zachary and S. Torquato, J. Stat. Mech. P12015 (2009).

\end{thebibliography}
\end{document}